\documentclass[aps,prd,reprint,groupedaddress,amsmath,amssymb,preprintnumbers,showpacs]{revtex4-1}
\usepackage{dcolumn}
\usepackage{graphicx}
\usepackage{subfigure}
\usepackage{color}
\usepackage[normalem]{ulem}

\begin{document}

\title{Newtonian versus relativistic cosmology}
\author{Samuel F.~Flender}
\email{samuel.flender@helsinki.fi}
\affiliation{Physics Department, University of Helsinki and Helsinki Institute of Physics, P.O. Box 64, FI-00014 Helsinki, Finland.}
\author{Dominik J.~Schwarz}
\email{dschwarz@physik.uni-bielefeld.de}
\affiliation{Fakult{\"a}t f{\"u}r Physik, Universit{\"a}t Bielefeld, Postfach 100131, D-33501 Bielefeld, Germany.}

\date{\today}

\preprint{HIP-2012-15/TH, BI-TP-2012/26}

\begin{abstract}
We show how the relativistic matter and velocity power spectra behave in different gauges. We construct a new gauge where both spectra coincide with Newtonian theory on all scales. However, in this gauge there are geometric quantities present which do not exist in Newtonian theory, for example the local variation of the Hubble parameter. Comparing this quantity to second order Newtonian quantities, we find that Newtonian theory is inaccurate on scales larger than 10 Mpc. This stresses the importance of relativistic corrections to Newtonian cosmological N-body simulations on these scales.
\end{abstract}


\maketitle

\section{Introduction}

At early times the Universe was very close to isotropic and homogeneous, the best indication being the tiny fluctuations (about $\sim10^{-5}$) in the cosmic microwave background (CMB) radiation \cite{Bennett:1996}. However, the structure of the Universe that we observe today is very inhomogeneous on scales below the homogeneity scale of $\sim100\,{\rm Mpc}$ \cite{Hogg:2004vw,Scrimgeour:2012wt}. There are galaxies (with mass collection radius up to $\sim1\,{\rm Mpc}$) and clusters of galaxies (at scales of $\sim10\,{\rm Mpc}$). Up to scales of $\sim100\,{\rm Mpc}$ we find superclusters and large voids \cite{Gott:2003}.

The theory of structure formation connects the early homogeneous Universe with the inhomogeneous Universe we observe today. It is assumed that there exist initially small density perturbations, which grow due to gravitational attraction. The mathematical tool is relativistic cosmological perturbation theory  \cite{Mukhanov:1990}, which considers small perturbations on a homogeneous and isotropic Friedmann-Robertson-Walker (FRW) space-time. However, in this theory there are subtleties arising due to the freedom of gauge, i.e.~choosing the correspondence between perturbed and background quantities \cite{Bardeen:1980,Bardeen:1988}.

Another approach is Newtonian cosmology, see e.g.~\cite{Bertschinger:1993}. One has to keep in mind that Newtonian theory is wrong in that it assumes instantaneous gravitational interaction and infinite speed of light. However, because of its simplicity compared to the relativistic theory, the Newtonian equations are the preferred choice in cosmological N-body simulations, which are used to test our understanding of structure formation.

But the question is, how reliable are these simulations, since they use Newtonian rather than relativistic equations. Here we ask, how reliable is Newtonian cosmology on large scales?

In order to answer this question, people have so far followed mainly two different strategies. One strategy is to choose a gauge and then calculate the relativistic corrections appearing in this gauge. It turns out that in any gauge that is not the synchronous gauge there are relativistic corrections appearing to the density contrast on large scales, see e.g. \cite{Zhang:2011,Dent:2008}. 

The other strategy is based on the use of gauge-invariant variables. Gauge-invariant quantities were first introduced by Bardeen in \cite{Bardeen:1980} (see also \cite{Mukhanov:1990} for a more recent review) and are the basis for the work of Hwang and Noh, who found an exact correspondence up to second order between Newtonian and relativistic cosmology \cite{Hwang:2005, Hwang:2012}. 

The relativistic-Newtonian correspondence up to second order found in \cite{Hwang:2005} holds in the following way: The Newtonian density contrast is identified with a relativistic gauge-invariant expression that reduces to the density contrast on comoving hypersurfaces. However, the Newtonian gravitational and velocity-potentials are identified with gauge-invariant expressions which reduce to  the relativistic potentials on zero-shear hypersurfaces. Using the same strategy, Hwang and Noh also found a correspondence between relativistic cosmological perturbation theory and the first order post-Newtonian approximation (1PN) \cite{Noh:2012su}. Again, this correspondence holds if one identifies the 1PN perturbation fields with relativistic gauge-invariant quantities that are interpreted on different hypersurfaces. 

However, one needs to keep in mind that any physical problem is described by a set of equations of motion and appropriate initial conditions. The initial conditions must be specified on a spatial Cauchy hypersurface, which in the context of cosmological perturbation theory corresponds to a particular foliation of space-time, i.e. to a hypothetical observer who is able to determine physical quantities on a spatial hypersurface. The relativistic-Newtonian correspondence mixes the quantities defined on different spatial hypersurfaces and thus no hypothetical observer in the Einsteinian world could actually determine these combined quantities.

Another recent approach from Green and Wald uses a whole new framework to investigate the problem \cite{Green:2010, Green:2011}.  Using their framework, the authors construct a dictionary between Newtonian solutions and relativistic solutions. However, in order to reconstruct the exact full relativistic structure out of the Newtonian solution, additional differential equations need to be solved (e.g. eq. (2.44) in ref. \cite{Green:2011}). A similar dictionary is proposed by Chisari and Zaldarriaga \cite{Chisari:2011}. They show how relativistic corrections can be absorbed into initial data of simulations and how to modify the Newtonian coordinates to be able to compare them to relativistic coordinates. Both strategies seem to result in extra input and extra numerical work on top of the Newtonian calculation.

In this work we give a \emph{quantitative} estimate of when the relativistic corrections to the Newtonian solutions are important. In order to do so, we first construct a gauge in which both the density contrast and the velocity perturbations coincide with Newtonian theory. Then, the significance of the relativistic corrections to Newtonian theory can be evaluated by considering the local modification of the Hubble parameter, which appears in this gauge, and comparing it to second order Newtonian quantities.

Let us note that the question of this work is different from the question what a real observer can see in the Universe. The latter has been recently discussed in \cite{Yoo:2009au, Yoo:2010, LopezHonorez:2011cy, Bonvin:2011bg}.

Throughout this work we restrict our attention to an Einstein-de Sitter background cosmology, because it provides an accurate description for most of the history of the Universe and has simple expressions for the scale factor ($a \propto \tau^2$, where $\tau$ denotes conformal time) and the conformal Hubble parameter ($\mathcal{H} = 2/\tau$). We adopt a Hubble constant of $H_0 = 71\,\rm{km/s/Mpc}$, in agreement with measurements of the Wilkinson Microwave Anisotropy Probe (WMAP) \cite{Larson:2010gs}.

The outline is as follows: In section II we review Newtonian cosmological perturbation theory and present the equations up to second order. In section III we describe relativistic cosmological perturbation theory and give an overview over solutions in the six most common gauges. In particular, we show how the shapes of power spectra vary from gauge to gauge. We find that none of these gauges gives both matter- and velocity spectra in correspondence with Newtonian theory. In section IV we introduce a new gauge with the property that it coincides with Newtonian theory on all scales regarding the density contrast \emph{and} the velocity perturbation, the Newtonian matter gauge. Using the strategy described above, we conclude that relativistic corrections are more important than second order Newtonian effects on scales larger than 10 Mpc.

\section{Newtonian cosmological perturbation theory}
\subsection*{First order}

Newtonian cosmological perturbation theory results in perturbed Newtonian equations (continuity equation, Euler equation and Poisson equation) on an expanding FRW background. In this model the linear equations can be written in Fourier space \cite{Bertschinger:1993}:
\begin{eqnarray}
\delta^{(1)\prime}+i\mathbf{k}\cdot\mathbf{v}^{(1)} & = & 0,\\
\mathbf{v}^{(1)\prime}+\mathcal{H}\mathbf{v}^{(1)}  & = & -i\mathbf{k}\phi^{(1)}, \\
-k^{2}\phi^{(1)} & = & \frac{3}{2}\mathcal{H}^{2}\delta^{(1)},
\end{eqnarray}
where $\delta^{(1)}$ is the first order density contrast, $\mathbf{v}^{(1)}$ is the first order velocity perturbation, $\phi^{(1)}$ is the first order gravitational potential and $k$ is the comoving wavenumber. The linear equations can be decoupled into scalar and vector parts. It turns out that in an Einstein-de Sitter model the vector contribution to $\mathbf{v}^{(1)}$ is decaying and hence can be neglected (see e.g.~\cite{Flender:2012}). 

The remaining growing solutions for the perturbations are:
\begin{eqnarray}
\phi^{(1)}   & = & \phi^{(1)}(\mathbf{k}),\\
\delta^{(1)} & = & -\frac{2}{3}\frac{k^{2}}{\mathcal{H}^{2}}\phi^{(1)},\\
\mathbf{v}^{(1)}   & = & -\frac{2}{3}\frac{i\mathbf{k}}{\mathcal{H}}\phi^{(1)}.
\end{eqnarray}

We assume a Harrison-Zel'dovich spectrum for the primordial curvature perturbation field, with an amplitude $A=4.88\cdot10^{-5}$, which corresponds to measurements of WMAP \cite{Larson:2010gs}. It follows that the initial condition for $\phi^{(1)}$ can be written as (see e.g. \cite{Flender:2012}):
\begin{equation}
|\phi^{(1)}|=\frac{3}{5}A(2\pi^{2})^{1/2}k^{-3/2}T(k),\label{eq:initial condition}
\end{equation}
where $T(k)$ is the transfer function, which accounts for suppressed growth of structure during the radiation dominated epoch. We adopt the form of $T(k)$ from \cite{Bardeen:1986} for a pure cold dark matter model (i.e., without any baryonic matter).

\subsection*{Second order}

At second order the Newtonian equations are more complicated. For a detailed derivation, see \cite{Flender:2012}. In Fourier space they read
\begin{eqnarray}
\delta^{(2)\prime}+i\mathbf{k}\cdot\mathbf{v}^{(2)} & = & -\frac{i\mathbf{k}}{(2\pi)^{3}}\cdot(\delta^{(1)}\star
\mathbf{v}^{(1)}),\\
\mathbf{v}^{(2)\prime}+\mathcal{H}\mathbf{v}^{(2)}+i\mathbf{k}\phi^{(2)} & = & -\frac{i\mathbf{k}}{(2\pi)^{3}}
(\mathbf{v}^{(1)}\star\mathbf{v}^{(1)}), \\
k^{2}\phi^{(2)} + \frac{3}{2}\mathcal{H}^2\delta^{(2)} & = & 0,
\end{eqnarray}
where a star denotes a convolution. Note that the structure of the second order equations is very similar to that of the first order equations. However, there are additional source terms, consisting of convolutions of first order perturbations, which we have written on the r.h.s.~of each equation. Due to these source terms, the second order perturbations grow faster than the first order perturbations. Solutions can be obtained numerically. Plots of the first and second order density contrast are shown below in figure \ref{d1d2dk}, where we compare them to a relativistic quantity, the local variation of the Hubble parameter.

\section{Relativistic cosmological perturbation theory}

\begin{figure*}
\centering
\subfigure{\includegraphics[scale=0.66]{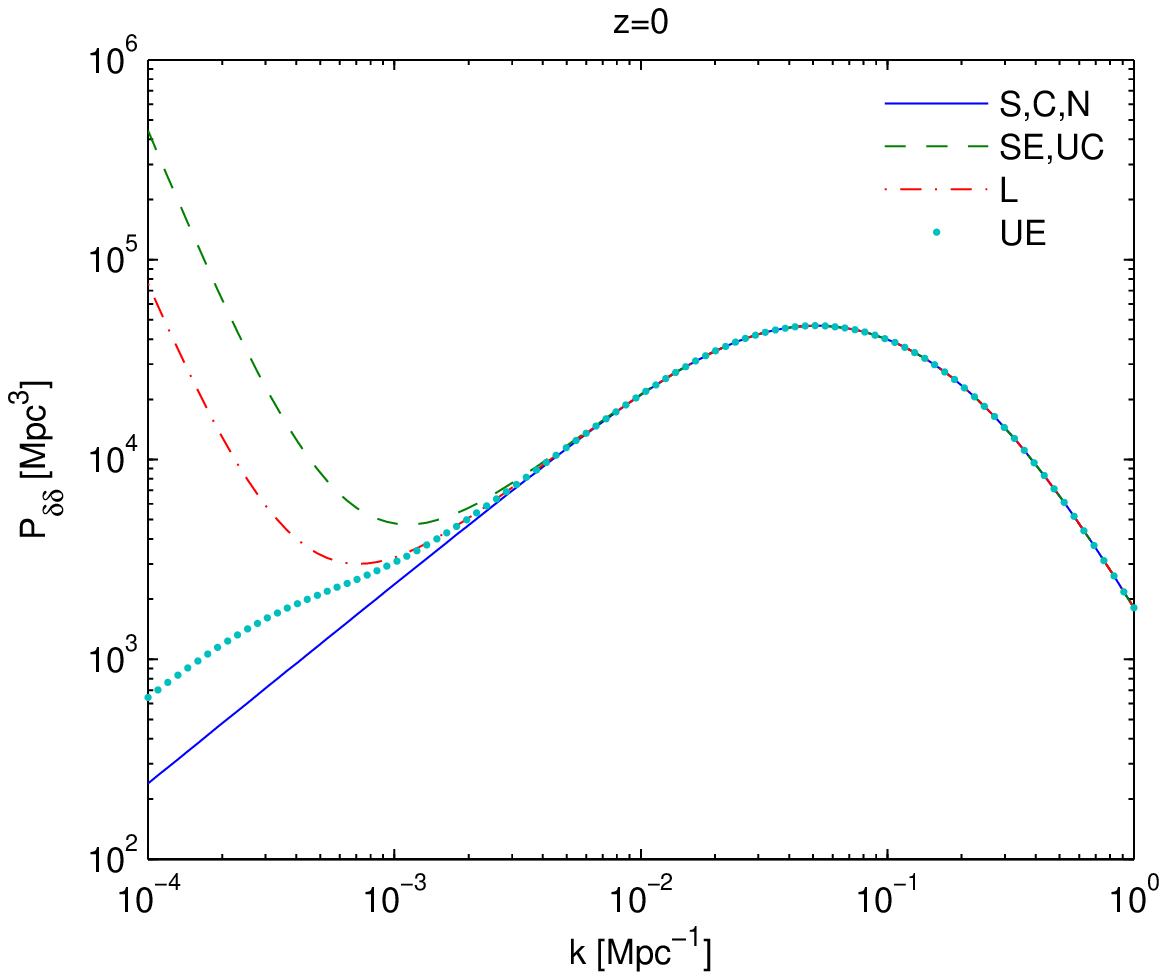}}
\subfigure{\includegraphics[scale=0.66]{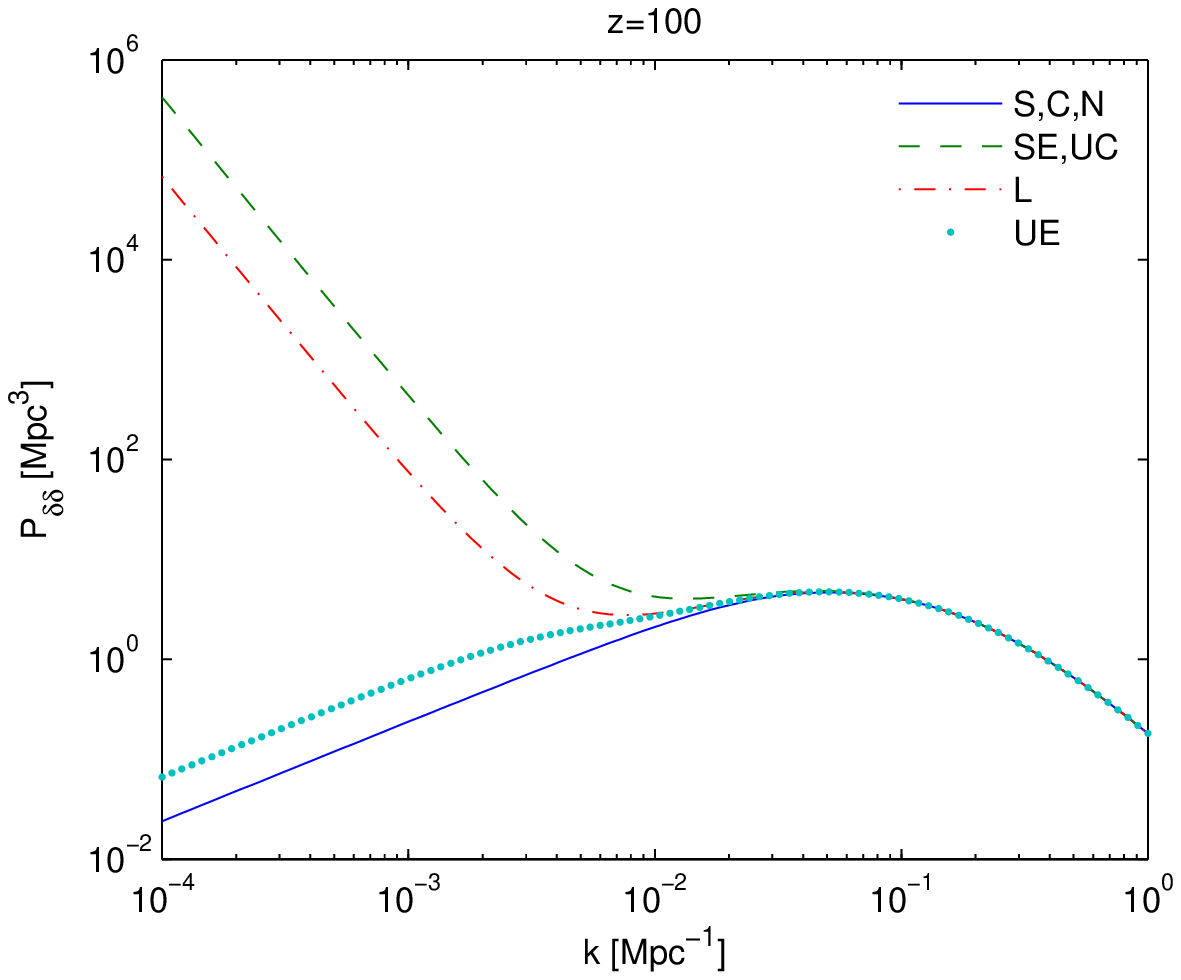}}\\
\subfigure{\includegraphics[scale=0.66]{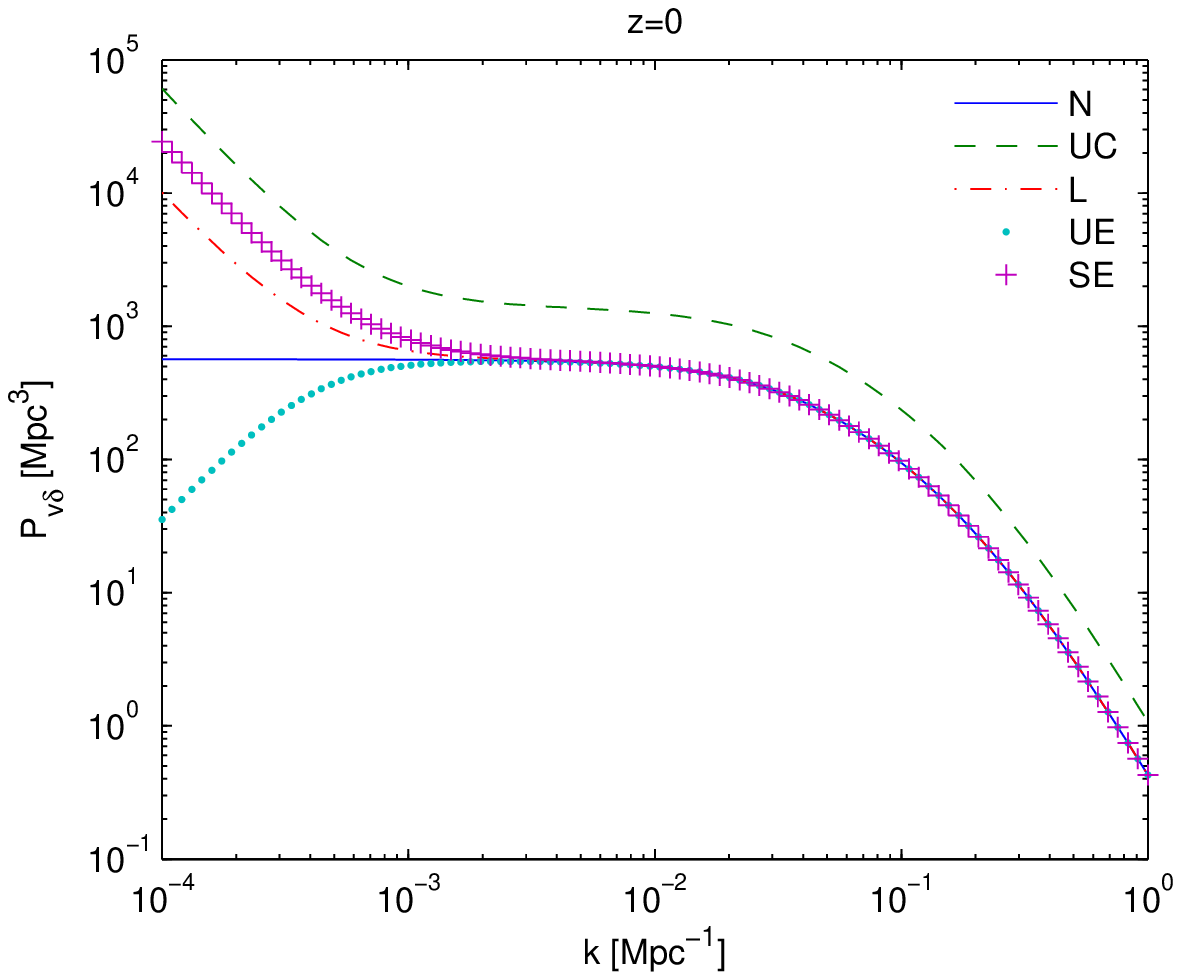}}
\subfigure{\includegraphics[scale=0.66]{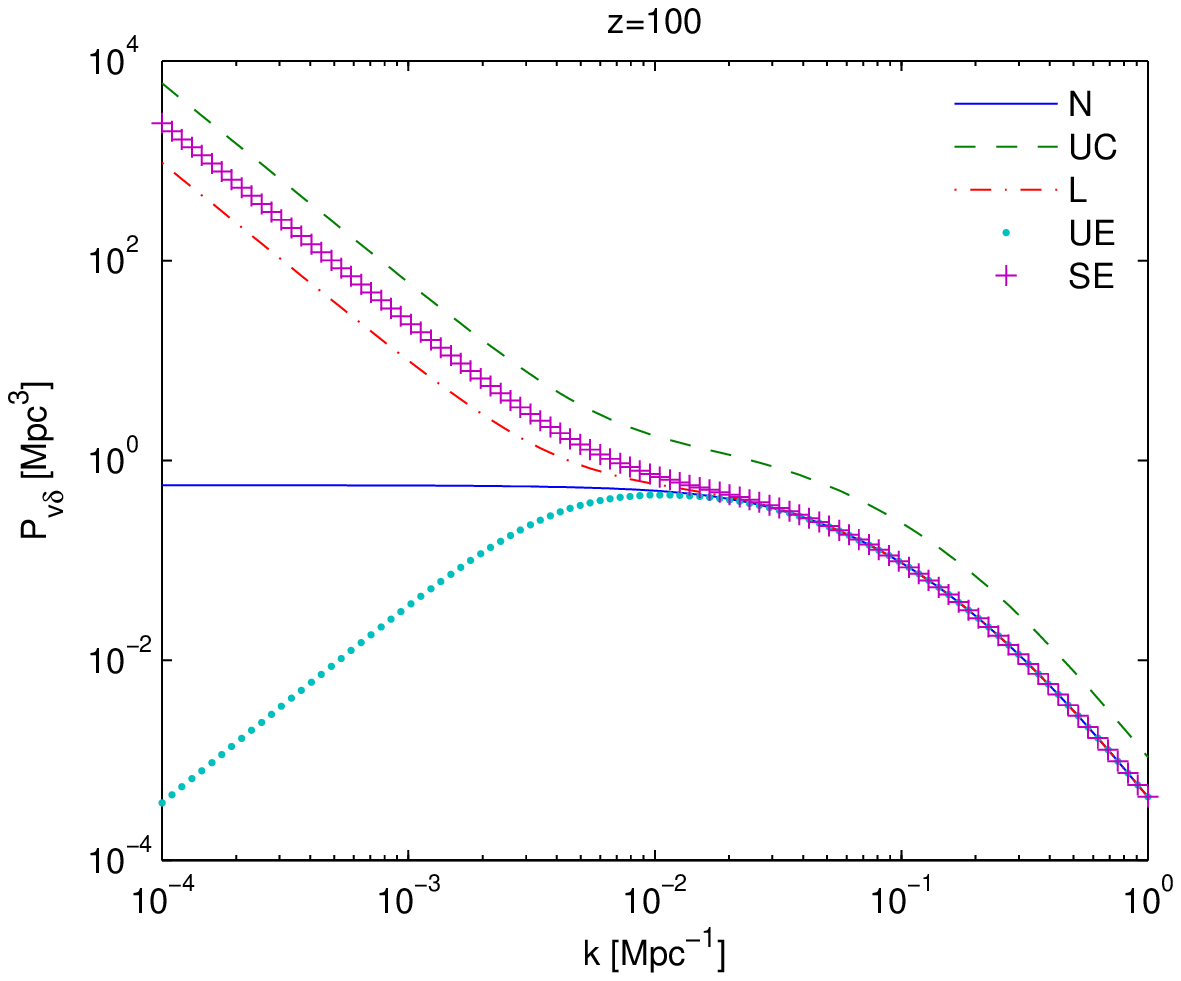}}\\
\subfigure{\includegraphics[scale=0.66]{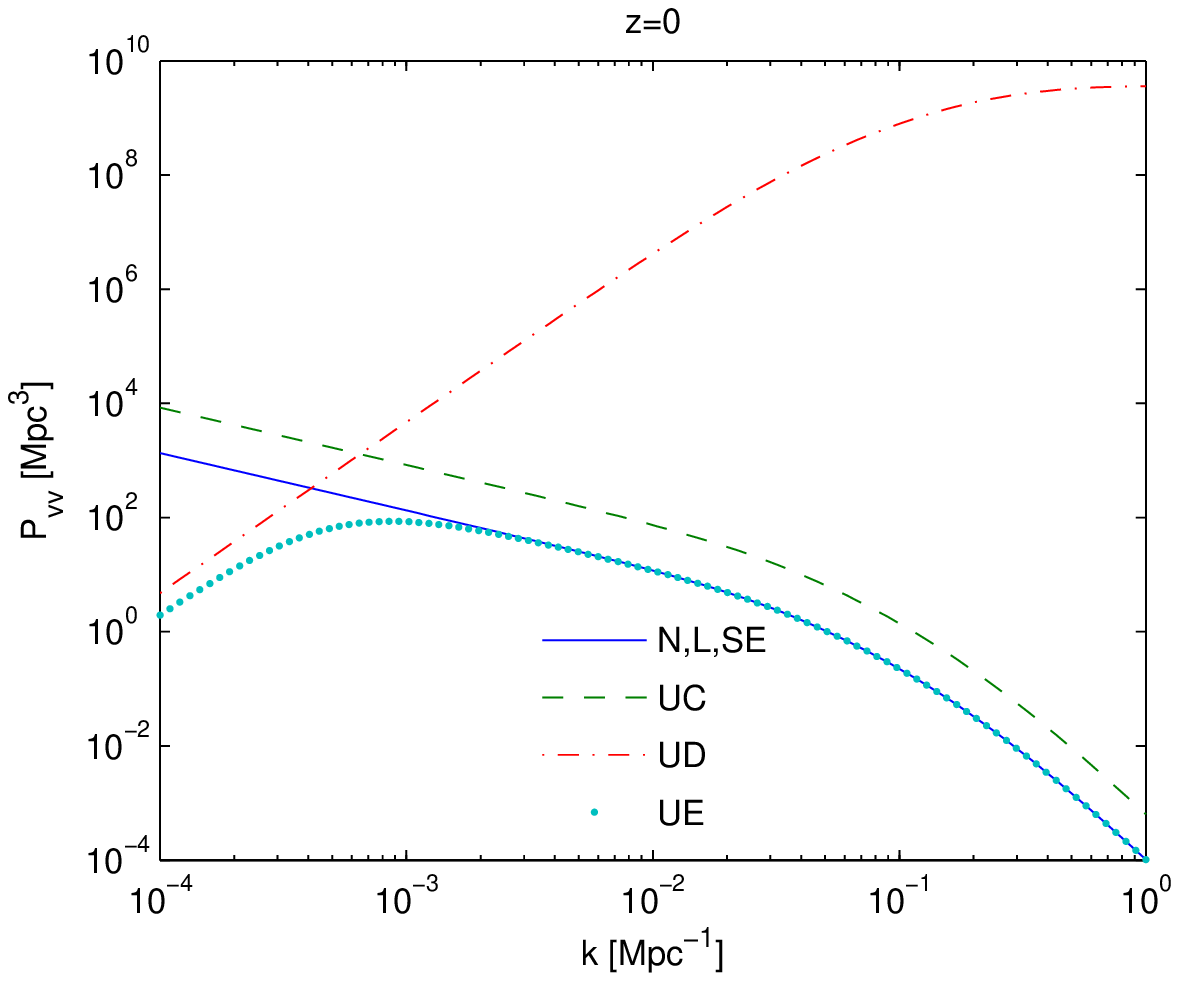}}
\subfigure{\includegraphics[scale=0.66]{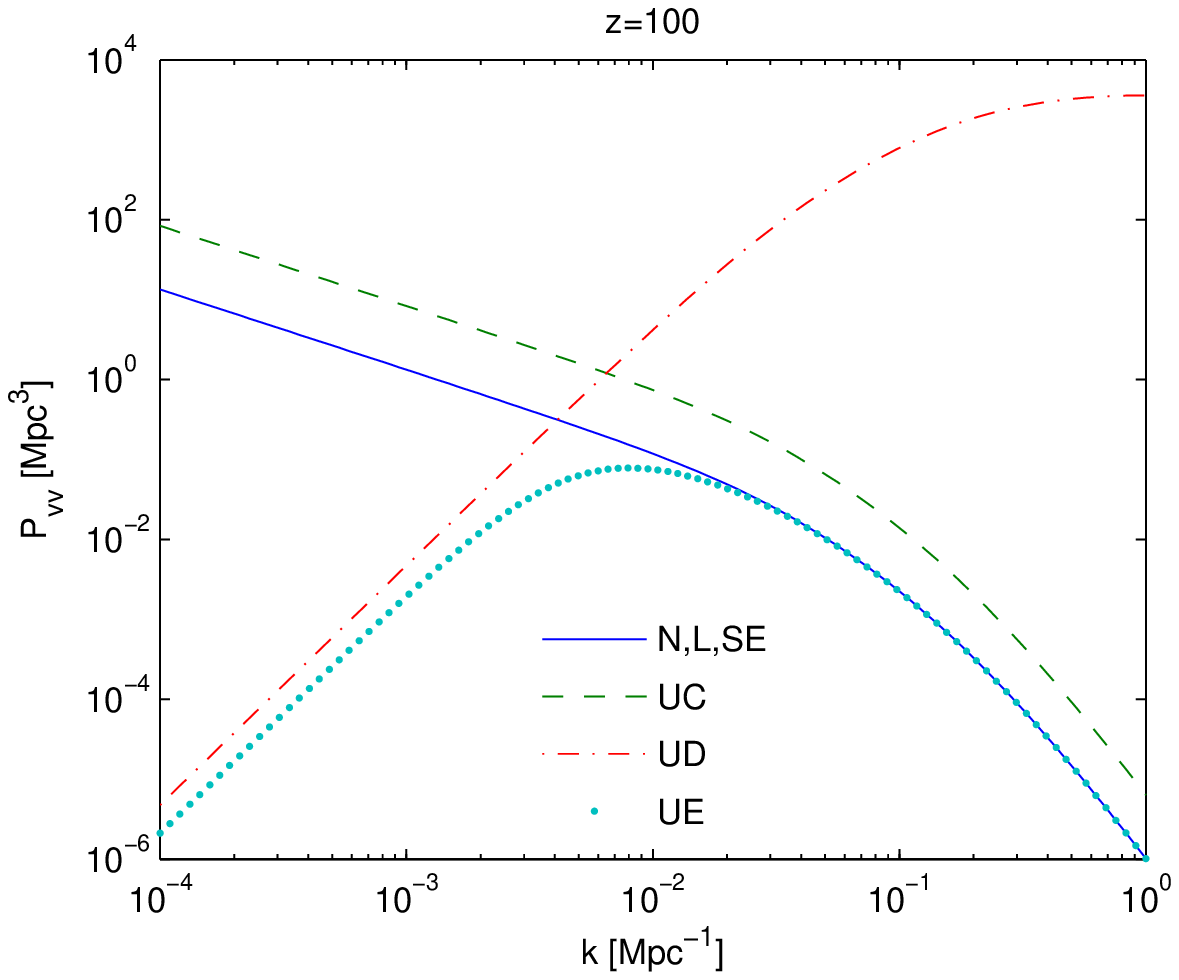}}
\caption{\label{spectra} 
Matter power spectrum ($P_{\delta\delta}\equiv|{\delta}|^{2}$), matter-velocity power spectrum 
($P_{\bf{v}\delta}\equiv|{\bf{v}}\delta|$) and velocity power spectrum 
($P_{\bf{vv}}\equiv|{\bf{v}}|^2$) according to relativistic perturbation theory in different gauges 
(defined in the text) 
and according to Newtonian theory (labelled by "N"), for $z=0$ (left panels) and $z=100$ (right panels).}
\end{figure*}

\begin{table*}
\begin{ruledtabular}
\begin{tabular}{ccccccc}
Gauge & L & UC & SE & S/C & UD & UE\\
\hline
$\phi$ & $\Phi$ & $\frac{5}{2}\Phi$ & $\frac{5}{2}\Phi$ & 0 & $-\frac{5}{9}\frac{k^{2}}{\mathcal{H}^{2}}\Phi$ & $\Phi-\frac{1}{a}(a\frac{3\mathcal{H}\Phi}{k^{2}+\frac{9}{2}\mathcal{H}^{2}})^{\prime}$
\tabularnewline
$\psi$ & $\Phi$ & 0 & 0 & $\frac{5}{3}\Phi$ & $\frac{5}{3}\Phi+\frac{2}{9}\frac{k^{2}}{\mathcal{H}^{2}}\Phi$ & $\Phi+\frac{3\mathcal{H}^{2}\Phi}{k^{2}+\frac{9}{2}\mathcal{H}^{2}}$
\tabularnewline
$w$ & 0 & 0 & $-\frac{\Phi}{\mathcal{H}}$ & 0 & 0 & 0
\tabularnewline
$h$ & 0 & $\frac{\Phi}{\mathcal{H}^{2}}$ & 0 & $-\frac{2}{3\mathcal{H}^{2}}\Phi$ & $-\frac{2}{3\mathcal{H}^{2}}\Phi-\frac{1}{3}\frac{k^{2}}{\mathcal{H}^{4}}\Phi$ & $-\int^{\tau}d\tau^{\prime}\frac{3\mathcal{H}\Phi}{k^{2}+\frac{9}{2}\mathcal{H}^{2}}$
\tabularnewline
\hline 
$v$ & -$\frac{2}{3}\frac{1}{\mathcal{H}}\Phi$ & -$\frac{5}{3}\frac{1}{\mathcal{H}}\Phi$ & -$\frac{2}{3}\frac{1}{\mathcal{H}}\Phi$ & 0 & $\frac{2}{9}\frac{k^{2}}{\mathcal{H}^{3}}\Phi$ & -$\frac{2}{3}\frac{1}{\mathcal{H}}\Phi+\frac{3\mathcal{H}\Phi}{k^{2}+\frac{9}{2}\mathcal{H}^{2}}$
\tabularnewline
$\delta$ & $-(2+\frac{2}{3}\frac{k^{2}}{\mathcal{H}^{2}})\Phi$ & $-(5+\frac{2}{3}\frac{k^{2}}{\mathcal{H}^{2}})\Phi$ & $-(5+\frac{2}{3}\frac{k^{2}}{\mathcal{H}^{2}})\Phi$ & $-\frac{2}{3}\frac{k^{2}}{\mathcal{H}^{2}}\Phi$ & 0 & $-(2+\frac{2}{3}\frac{k^{2}}{\mathcal{H}^{2}})\Phi+\frac{9\mathcal{H}^{2}\Phi}{k^{2}+\frac{9}{2}\mathcal{H}^{2}}$
\tabularnewline
\hline 
$^{(3)}R$ & $-4\frac{k^{2}}{a^{2}}\Phi$ & 0 & 0 & $-\frac{20}{3}\frac{k^{2}}{a^{2}}\Phi$ & $\frac{20}{9}\frac{k^{4}}{a^{2}\mathcal{H}^{2}}\Phi$ & $-4\frac{k^{2}}{a^{2}}(\Phi+\frac{3\mathcal{H}^{2}\Phi}{k^{2}+\frac{9}{2}\mathcal{H}^{2}})$
\tabularnewline
$\chi$ & 0 & $a\frac{\Phi}{\mathcal{H}}$ & $a\frac{\Phi}{\mathcal{H}}$ & $-\frac{2}{3}a\frac{\Phi}{\mathcal{H}}$ & $-\frac{2}{3}a\frac{\Phi}{\mathcal{H}}(1+\frac{k^{2}}{\mathcal{H}^{2}})$ & $-\frac{3a\mathcal{H}\Phi}{k^{2}+\frac{9}{2}\mathcal{H}^{2}}$
\tabularnewline
$\kappa$ & $3\frac{\mathcal{H}\Phi}{a}$ & $(\frac{15}{2}+\frac{k^{2}}{\mathcal{H}^{2}})\frac{\mathcal{H}\Phi}{a}$ & $(\frac{15}{2}+\frac{k^{2}}{\mathcal{H}^{2}})\frac{\mathcal{H}\Phi}{a}$ & $(\frac{15}{2}-\frac{2}{3}\frac{k^{2}}{\mathcal{H}^{2}})\frac{\mathcal{H}\Phi}{a}$ & $-\frac{2}{3}\frac{k^{2}}{\mathcal{H}^{2}}(5+\frac{k^{2}}{\mathcal{H}^{2}})\frac{\mathcal{H}\Phi}{a}$ & 0\tabularnewline
\end{tabular}
\end{ruledtabular}
\caption{\label{tab:gauges} Overview - growing modes of first order scalar perturbations for dust models according to relativistic cosmological perturbation theory in longitudinal (L), uniform curvature (UC), spatially Euclidean (SE), synchronous (S), comoving (C), uniform density (UD), and uniform expansion (UE) gauge. The quantities are arranged in the following way: The upper group contains perturbations in the metric tensor, the middle group contains perturbations in the energy-momentum tensor (peculiar velocity potential and density contrast) and the lower group contains derived geometrical quantities (intrinsic curvature, shear, and expansion rate). All solutions are expressed as function of Bardeen's metric potential $\Phi$.}
\end{table*}

In relativistic cosmological perturbation theory we consider small perturbations of the homogeneous and isotropic metric and the energy-momentum tensor of a perfect fluid. The evolution of these perturbations follows from the covariant conservation of  the energy-momentum tensor and Einstein's field equations. We define the perturbed metric
\begin{equation}
g_{\mu\nu}=a^{2}\begin{pmatrix}\begin{array}{cc}
-1-2\phi & w_{;i}\\
w_{;i} & (1-2\psi)\delta_{ij}+2h_{;ij}
\end{array}\end{pmatrix}.
\end{equation}
We restrict our attention to scalar perturbations at linear order. There are hence 4 degrees of freedom in the metric perturbation, corresponding to $\phi,$ $\psi$, $w$ and $h$. However, two of them can be fixed due to the gauge freedom. This corresponds to choosing $\xi^{0}$ and $\xi$ in the gauge transformation 
\begin{equation}
\begin{pmatrix}\tau\\
\mathbf{x}
\end{pmatrix}\rightarrow\begin{pmatrix}\tau+\xi^{0}\\
\mathbf{x}+\nabla\xi
\end{pmatrix},
\end{equation}
where $\tau$ is the conformal time and $\mathbf{x}$ are comoving coordinates. In the following we will discuss the advantages of choosing $w=0$ as a gauge condition. The proper time between two events along a worldline is given by
\begin{widetext}
\begin{equation}
\int\sqrt{-ds^{2}}=\int d\tau a\sqrt{1+2\phi-2w_{;i}\frac{dx^{i}}{d\tau}-[(1-2\psi)\delta_{ij}+2h_{;ij}]\frac{dx^{i}}{d\tau}\frac{dx^{j}}{d\tau}}.
\end{equation}
\end{widetext}
Now we expand the integrand and keep only terms up to $\mathcal{O}(gv_{i})$, where $g\in\{\phi,\psi,w,h,v_{i}\}$, which gives 
\begin{equation}
\int\sqrt{-ds^{2}}=\int d\tau a\left(1+\phi-\mathbf{v}\cdot\nabla w-\frac{\mathbf{v}^{2}}{2}\right).
\end{equation}
The proper time is extremal if and only if the Lagrangian $L=a(1+\phi-\mathbf{v}\cdot\nabla w-\frac{\mathbf{v}^{2}}{2})$ satisfies the \emph{Euler-Lagrange equation}
\begin{equation}
\frac{d}{d\tau}\frac{\partial L}{\partial\mathbf{v}}=\frac{\partial L}{\partial\mathbf{x}}.
\end{equation}
This gives in first order
\begin{equation}
\label{ELGl}
\frac{1}{a}\frac{d[a(\mathbf{v}+\nabla w)]}{d\tau}=-\nabla\phi.
\end{equation}

It is convenient to choose $w=0$, so that the Euler-Lagrange equation coincides exactly with Newton's law for the motion of particles in an expanding universe,
\begin{equation}
\label{NewtonLaw}
\frac{1}{a}\frac{d(a\mathbf{v})}{d\tau}=-\nabla\phi.
\end{equation}
Thus, in a gauge with $w=0$ we can identify the metric perturbation $\phi$ with the Newtonian gravitational potential. This is the reason why it is common to identify the Bardeen potential $\Phi$, which is the same as $\phi$ in the longitudinal gauge (where $w=h=0$), as the gravitational potential (the exact definition of $\Phi$ is  given below). In physical terms, $w=0$ implies that comoving observers do not change their coordinates between constant-time hypersurfaces.

After choosing $w=0$ there is still another degree of freedom, which can be fixed by a second gauge condition, e.g.

\begin{itemize}

\item \emph{Uniform curvature gauge} (UC). In this gauge we set $w=\psi=0$. It follows that the intrinsic Ricci curvature, which is given by $^{(3)}R=\frac{4}{a^{2}}\Delta\psi$, vanishes in this gauge \cite{Bardeen:1988}.

\item \emph{Longitudinal gauge} (L). In this gauge we set $w=h=0.$ It is equivalent to the \emph{zero shear gauge, }defined by $w=\chi=0$, where 
\begin{equation}
\chi\equiv a(h^{\prime}-w)
\end{equation}
generates the traceless part of the extrinsic curvature tensor and can be interpreted as shear in the normal worldlines \cite{Bardeen:1988}. Another common name for this gauge is \emph{conformal Newtonian gauge}. However, we will not use this name since it is not the only gauge where the particles move according to Newton's law of motion, as we have shown above. Bertschinger's \emph{Poisson gauge} reduces to the longitudinal gauge in the scalar sector \cite{Bertschinger:1993}. 

The longitudinal gauge is equivalent to the use of the \emph{Bardeen potentials} \citep{Bardeen:1980},
\begin{eqnarray}
\Phi & \equiv & \phi+\frac{1}{a}\left[(w-h')a\right]',\\
\Psi & \equiv & \psi-\mathcal{H}(w-h').
\end{eqnarray}

\item \emph{Uniform expansion gauge} (UE). In this gauge we set $w=\kappa=0$, where 
\begin{equation}
\kappa\equiv\frac{3}{a}(\psi^{\prime}+\mathcal{H}\phi)-\frac{1}{a^{2}}\Delta\chi
\end{equation}
is the perturbation of the trace of the extrinsic curvature tensor \cite{Bardeen:1988}. Since the full trace is given by $K=-3H+\kappa$, where $H$ is the Hubble expansion rate, we can identify $\delta H \equiv -{\kappa}/3$ as the perturbation in the expansion rate.

\item \emph{Uniform density gauge} (UD). In this gauge we set $w=\delta=0$. There are hence no density perturbations.

\item \emph{Synchronous gauge} (S). In this gauge we set $w=\phi=0$. It follows that observers at different places have synchronous clocks, since $\phi$ affects the time-time component of the metric tensor. There is a residual gauge freedom, which can easily be fixed by specifying that the peculiar velocity vanishes at a particular moment of time (see e.g.~\cite{Mukhanov:1990}.) For a matter dominated universe this coincides with the comoving gauge (see below).

\item \emph{Comoving gauge} (C). In this gauge we set $w=v=0$, where $v$ is the velocity potential, so that $\mathbf{v=\nabla}v$. There are hence no peculiar velocities in this gauge. It turns out that during matter domination the comoving gauge and the synchronous gauge coincide \cite{Flender:2012}. Hence, we will treat them as one gauge in the following.

\item An interesting gauge with $w\ne0$ is the \emph{spatially Euclidean gauge} (SE). Here we set $h=\psi=0$, so that the spatial part of the metric has Euclidean geometry.

\end{itemize}

The relativistic equations of motion for the fluctuations follow from the perturbed energy-momentum conservation equation, $T_{\nu;\mu}^{\mu}=0$, and the Einstein equation, $G_{\mu\nu}=8\pi GT_{\mu\nu}$, where $G_{\mu\nu}$ denotes the Einstein tensor and $T_{\mu\nu}$ is the energy-momentum tensor. In longitudinal gauge, covariant conservation of energy and momentum gives for dust (see e.g.~\cite{Mukhanov:1990})
\begin{eqnarray}
\delta_{\rm{GI,L}}^{\prime}+\Delta v_{\rm{GI,L}}-3\Psi^{\prime} & = & 0,\\
\nabla v_{\rm{GI,L}}^{\prime}+\mathcal{H}\nabla v_{\rm{GI,L}} & = & -\nabla\Phi,
\end{eqnarray}
and the Einstein equation gives
\begin{eqnarray}
3\mathcal{H}^{2}\Phi+3\mathcal{H}\Psi^{\prime}-\Delta\Psi & = & -\frac{3}{2}\mathcal{H}^2\delta_{\rm{GI,L}},\\
\Psi^{\prime}+\mathcal{H}\Phi & = & -\frac{3}{2}\mathcal{H}^2v_{\rm{GI,L}},\\
3\Psi^{\prime\prime}+3\mathcal{H}(\Phi^{\prime}+2\Psi^{\prime})+\Delta(\Psi-\Phi) & = & 0,\\
(\Psi-\Phi)_{,ij} & = & 0,
\end{eqnarray}
where we have introduced the gauge-invariant density contrast,
\begin{equation}
\delta_{\rm{GI,L}} \equiv \delta - 3\mathcal{H}(w-h^{\prime}),
\end{equation}
as well as the gauge-invariant velocity potential,
\begin{equation}
v_{\rm{GI,L}} \equiv v + h^{\prime}.
\end{equation}
Thus, the equations are written in gauge-invariant form. Note that the fields $\Phi,\,\delta_{\rm{GI,L}},\,v_{\rm{GI,L}}$ reduce simply to $\phi_{\rm{L}},\,\delta_{\rm{L}},\,v_{\rm{L}}$, respectively, in the longitudinal gauge. This tells us how to interpret this set of gauge-invariant fields: They measure the perturbations on zero-shear hypersurfaces.

In longitudinal gauge, the growing solutions are:
\begin{eqnarray}
\psi_{\rm L} = \phi_{\rm L} & = & \Phi = {\rm constant} ,\\
v_{\rm{L}} & = & -\frac{2}{3}\frac{\Phi}{\mathcal{H}},\\
\delta_{\rm{L}} & = & -2\Phi-\frac{2}{3}\frac{k^{2}}{\mathcal{H}}\Phi.
\end{eqnarray}

The solutions in other gauges can be obtained using the following gauge transformations \cite{Bardeen:1988}:
\begin{eqnarray}
\tilde{\phi} & = & \hat{\phi}-\mathcal{H}\xi^{0}-\xi^{0\prime},\\
\tilde{\psi} & = & \hat{\psi}+\mathcal{H}\xi^{0},\\
\tilde{w} & = & \hat{w}+\xi^{0}-\xi^{\prime},\\
\tilde{h} & = & \hat{h}-\xi,\\
\tilde{v} & = & \hat{v}+\xi^{\prime},\\
\tilde{\delta} & = & \hat{\delta}+3\mathcal{H}\xi^{0},\\
\tilde{\chi} & = & \hat{\chi}-a\xi^{0},\\
\tilde{\kappa} & = & \hat{\kappa}-\frac{9}{2}\mathcal{H}^{2}\frac{\xi^{0}}{a}+\Delta\frac{\xi^{0}}{a}.
\end{eqnarray}
The solutions in different gauges are given in table \ref{tab:gauges}.

Figure \ref{spectra} shows plots of the the matter power spectrum, $P_{\delta\delta}=|\delta|^{2}$, the matter-velocity power spectrum, $P_{\mathbf{v}\delta}=|\mathbf{v}\delta|$, and the velocity power spectrum, $P_{\mathbf{vv}}=|\mathbf{v}|^{2}$. Each spectrum is shown for different gauges and at two different redshifts: $z=100$, because this is where typical numerical simulations begin, and $z=0$, because this is where typical numerical simulations end. The initial condition for $\Phi$ is given by equation (\ref{eq:initial condition}). This is possible because we can identify $\Phi = \phi_{\rm L}$ as the Newtonian gravitational potential, as seen from (\ref{NewtonLaw}).  

Figure \ref{spectra} demonstrates that the relativistic matter power spectrum coincides with the Newtonian one on all scales in synchronous/comoving gauge, however there are no relativistic velocity perturbations. For the longitudinal gauge and the spatially Euclidean gauge the relativistic velocity power spectrum coincides with the Newtonian one on all scales, but the matter power spectra do not match on superhorizon scales. The other gauges give non-Newtonian results in both spectra. None of these gauges gives a correspondence to Newtonian theory both in the matter- and the velocity power spectrum on all scales.

Note that the claims about the Newtonian-relativistic correspondence found in literature are based on either the use of appropriate dictionaries \cite{Green:2011,Chisari:2011} or the appropriate mixing of gauge-invariant variables \cite{Hwang:2012}. We stress that the findings in this section are solutions as measured by one hypothetical observer, i.e. in one gauge. It seems to us that, in order to estimate the relativistic corrections to Newtonian cosmology, first of all we must seek a gauge in which both the density contrast and the matter velocity coincide.

\section{Newtonian matter gauge}

\begin{figure*}
\centering
\subfigure{\includegraphics[scale=0.66]{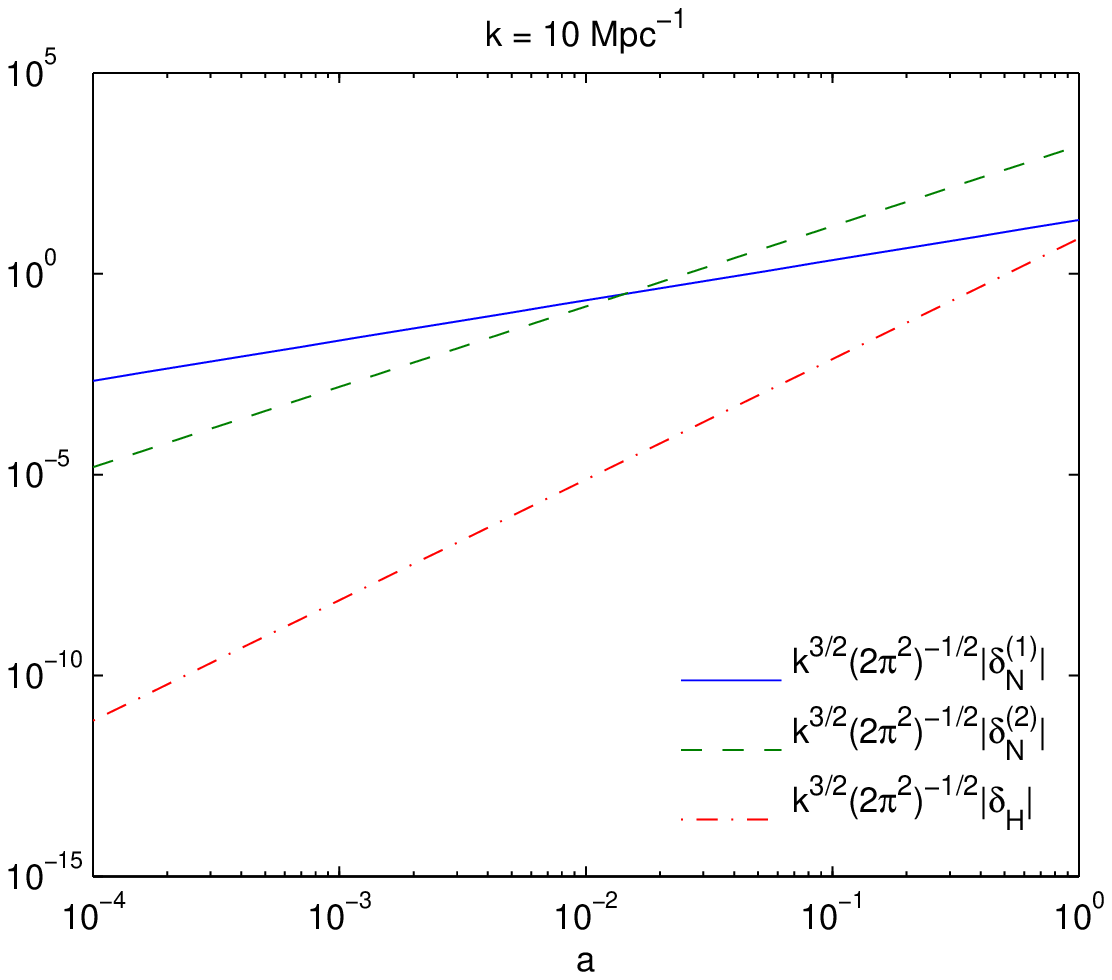}}\qquad
\subfigure{\includegraphics[scale=0.66]{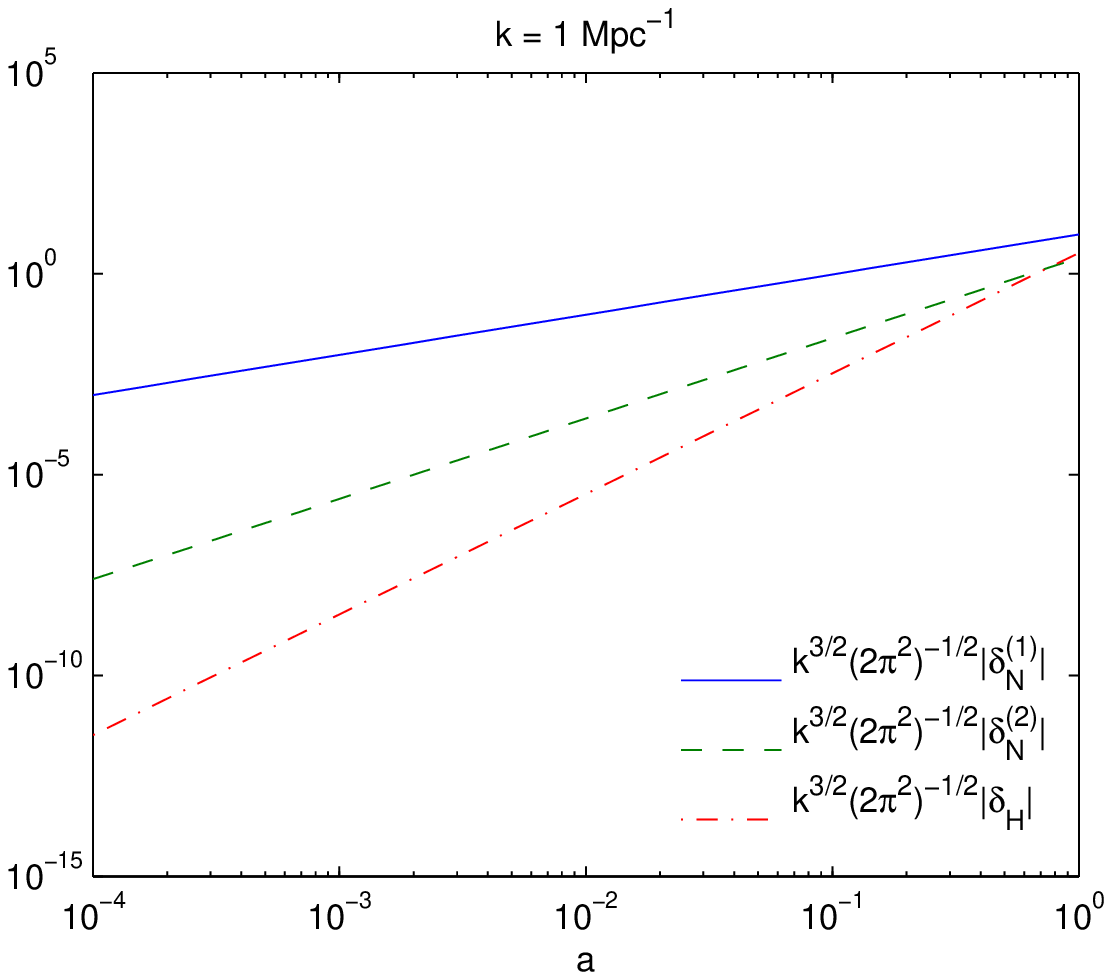}}\\
\subfigure{\includegraphics[scale=0.66]{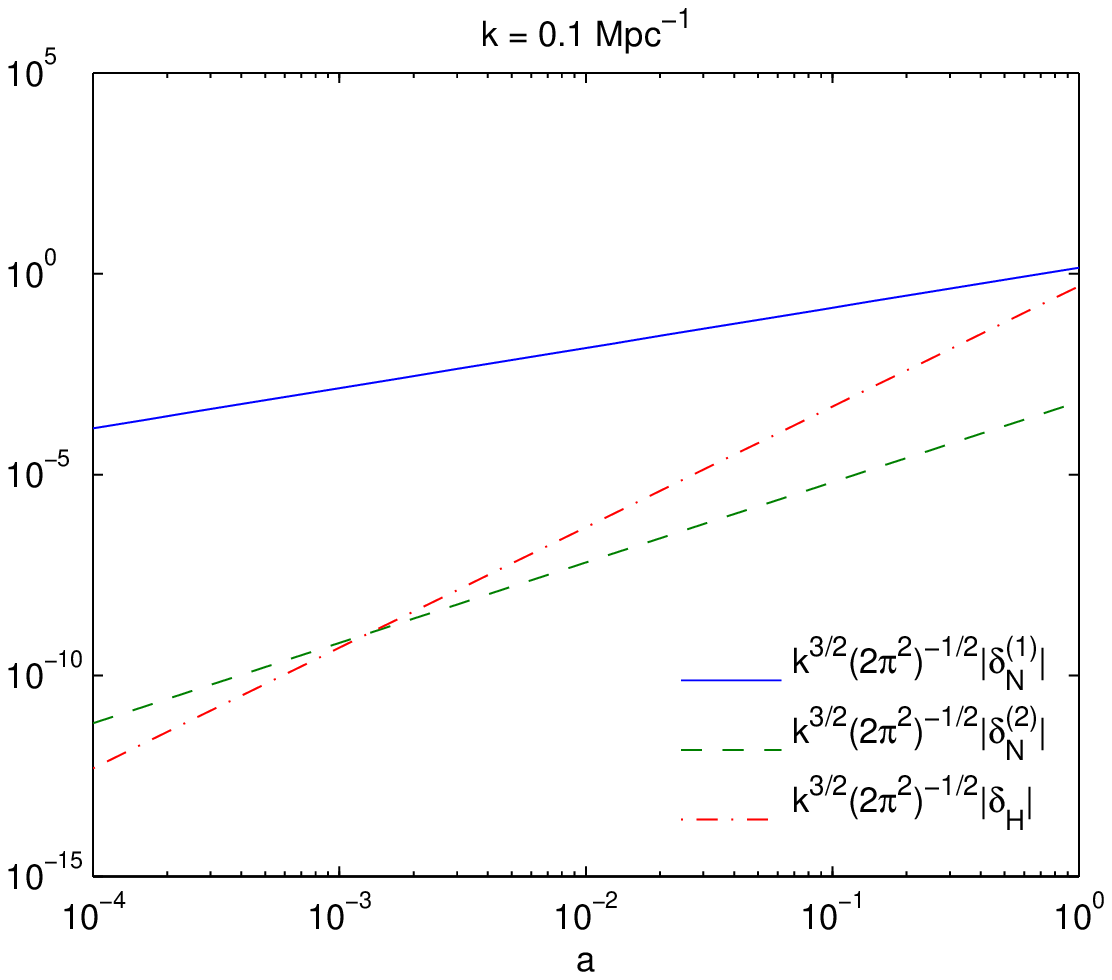}}\qquad
\subfigure{\includegraphics[scale=0.66]{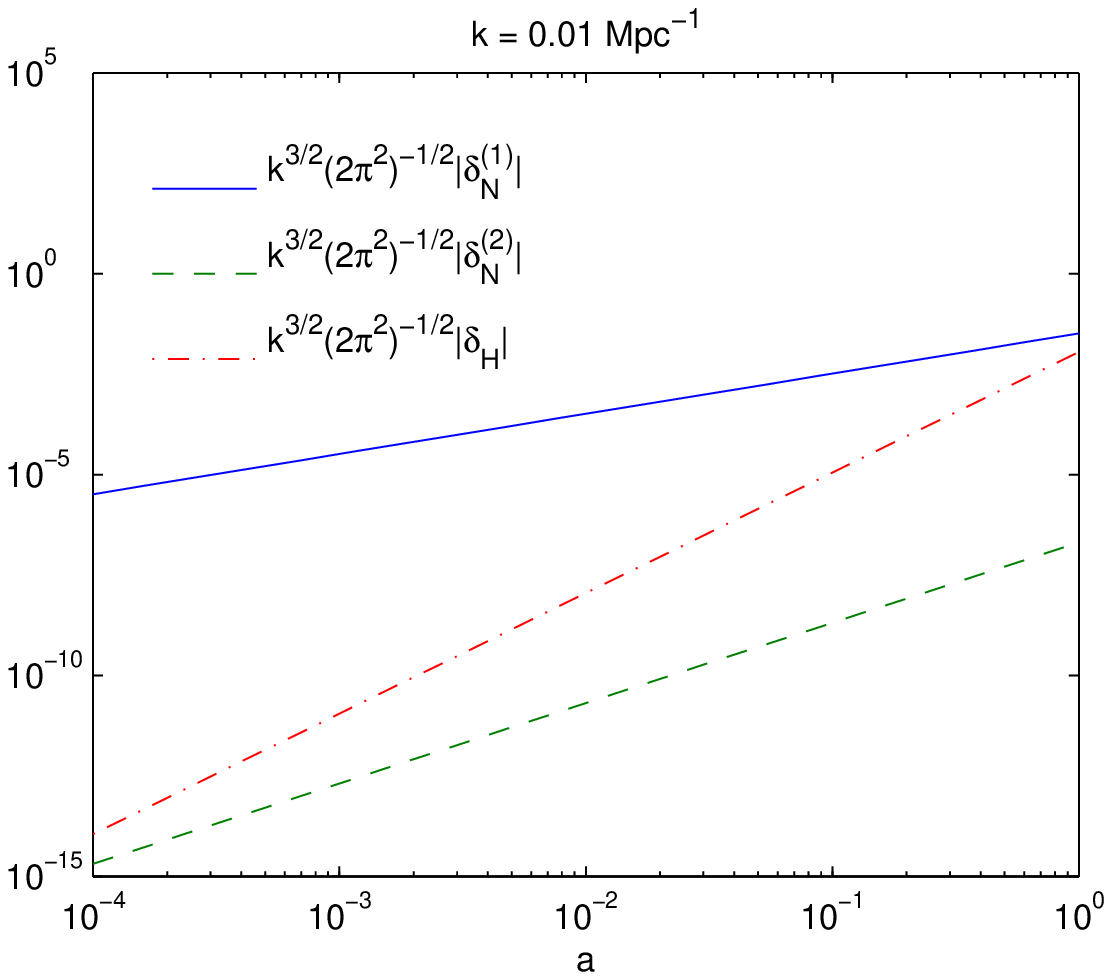}}
\caption{\label{d1d2dk} Density contrast in first and second order Newtonian cosmological perturbation theory, compared to the quantity $\delta_H$ (defined in the text), which represents the local modification of the Hubble rate and is a purely relativistic effect. The prefactor is chosen to make the quantities dimensionless. The different panels show the behaviour on different scales: on comoving length scales larger than 10 Mpc the relativistic modification is always larger than second order Newtonian effects, while on scales below 1 Mpc the relativistic modifications are small compared to second order Newtonian effects. These results hold for a spatially flat dust universe.}
\end{figure*}

\begin{figure}
\centering
\includegraphics[scale=0.66]{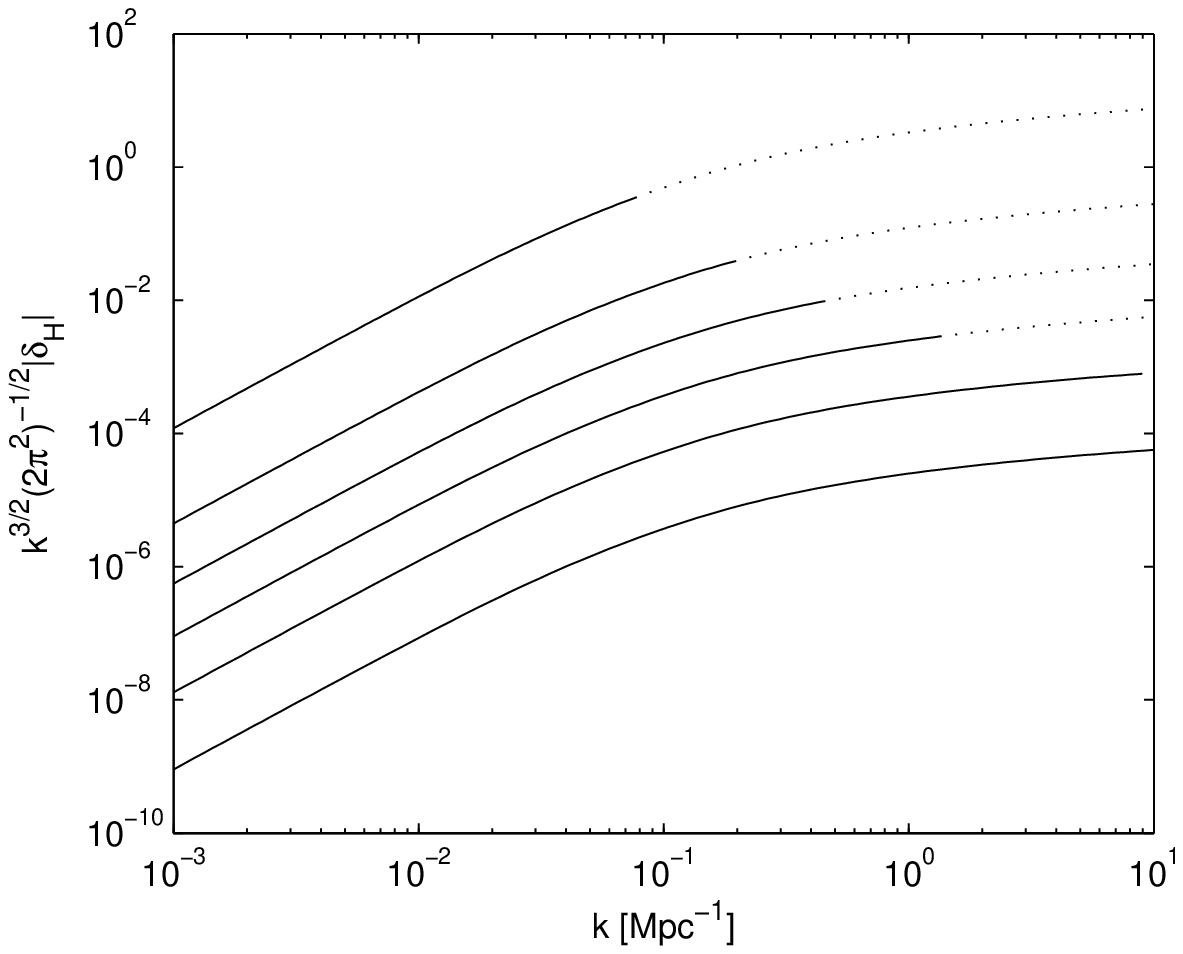}
\caption{\label{dhvsk} The local modification of the Hubble parameter, $\delta_H$, plotted against the comoving wavenumber $k$. From top to bottom, the lines denote growing redshifts of 0, 2, 5, 10, 20 and 50, respectively. Solid lines mark the regime of validity of linear perturbation theory. Our results do not strikly apply in the nonlinear regime (dotted lines).}
\end{figure}

By choosing $\xi^{0}$ and $\xi$ adequately, it is possible to construct a gauge in which both the density contrast and the peculiar velocity coincide with their Newtonian counterparts on all scales. We call this gauge the Newtonian matter gauge (NM), as all variables associated with the state of the matter (dust) agree with their Newtonian values. In order to construct it, we start in the longitudinal gauge, where the peculiar velocity already coincides with the Newtonian one. We then choose $\xi^{0}=\frac{2\Phi}{3\mathcal{H}}$ so that
\begin{equation}
\delta_{\rm N} = \delta_{\rm NM} = \delta_{\rm L}+3\mathcal{H}\xi^{0}.
\end{equation}
Note that the ($0,0$)-component of the metric perturbation vanishes in this gauge,
\begin{equation}
\phi_{\rm NM}=\Phi-\mathcal{H}\xi^{0}-\xi^{0\prime}=0,
\end{equation}
which is not a problem because the ($0,i$)-part of the metric perturbation is not zero in this gauge,
\begin{equation}
w_{\rm NM}=w_{\rm L}+\xi^{0}-\xi^{\prime}=\xi^{0}=\frac{2\Phi}{3\mathcal{H}},
\end{equation}
which means that we cannot identify $\phi_{\rm NM}$ as the Newtonian gravitational potential, as we discussed earlier. 

It is illuminating to evaluate (\ref{ELGl}) in the Newtonian matter gauge. Since $\phi_{\rm NM}=0$, we have
\begin{equation}
\frac{1}{a}\frac{d}{d\tau}[a(\mathbf{v}_{\rm NM}+\nabla w_{\rm NM})]=0.
\end{equation}
Using
\begin{equation}
\frac{1}{a}\frac{d}{d\tau}(a{w_{\rm NM}})
=\frac{1}{a}\frac{d}{d\tau}\left(a\frac{2\Phi}{3\mathcal{H}}\right)
=\frac{1}{\tau^2}\frac{d}{d\tau}\tau^3\frac{\Phi}{3}
=\Phi
\end{equation}
we recover Newton's equation of motion. This is not a surprise because we have chosen the gauge to give matter velocities in agreement with Newtonian theory.

In the Newtonian matter gauge we find a non-vanishing intrinsic curvature,
\begin{equation}
^{(3)}R_{\rm NM}=-\frac{20}{3}\frac{k^{2}}{a^{2}}\Phi,
\end{equation}
as well as a non-vanishing expansion rate perturbation,
\begin{equation}
-\kappa_{\rm NM}=\frac{2}{3}\frac{k^{2}}{a\mathcal{H}}\Phi,
\end{equation}
and non-vanishing shear, 
\begin{equation}
\chi_{\rm NM}=-\frac{2}{3}\frac{a}{\mathcal{H}}\Phi.
\end{equation}
All these quantities do not appear in Newtonian cosmology. In particular, we define the local modification of the Hubble expansion rate,
\begin{equation}
\delta_{H} \equiv \frac{(\delta H)_{\rm NM}}{H} = \frac{-\kappa_{\rm NM}}{3H}.
\end{equation}

Since the density contrast and the peculiar velocities coincide with Newtonian theory in this gauge, we can use $\delta_H$ to estimate the magnitude of relativistic modifications to the Newtonian theory. In Figure \ref{d1d2dk} we show the density contrast at first and second order Newtonian cosmological perturbation theory, as well as $\delta_H$ on different scales. We see that on comoving length scales smaller than 1 Mpc the second order Newtonian effects are always larger than the relativistic modifications. However, on comoving length scales larger than 10 Mpc the relativistic modification is more important than the second Newtonian order. This can have a drastic effect on the reliability of Newtonian N-body simulations of large cosmological volumes. Thus, Newtonian and general relativistic models of cosmology could coincide in the matter and velocity power spectra for all times, but would then disagree in their Hubble diagrams and thus in their conclusions on the expansion history of the Universe.

\begin{table}

\begin{tabular}{c|cc}
$k$ [Mpc$^{-1}$] & $z_{\rm NL}$ & $z_{\rm NN}$
\tabularnewline
\hline
0.01 & \emph{future} & $>10^4$ 
\tabularnewline
0.1 & 0.42 & 753
\tabularnewline
0.3 & 3.43 & 49.2
\tabularnewline
0.6 & 6.23 & 6.10
\tabularnewline
1 & 8.58 & 0.33
\tabularnewline
10 & 20.7 & \emph{future}
\end{tabular}

\caption{\label{anlann}$z_{\rm NL}$ and $z_{\rm NN}$ (defined in the text) for different comoving wavenumbers. For $k\gtrsim 0.6\,\rm{Mpc}^{-1}$ we have $z_{\rm NL}>z_{\rm NN}$, which means that these scales go nonlinear before relativistic effects become more important than second-order Newtonian effects. However, for $k\lesssim 0.6\,\rm{Mpc}^{-1}$ we find $z_{\rm NL}<z_{\rm NN}$, so that relativistic corrections become more important than the second order Newtonian terms before these scales go nonlinear.}
\end{table}

In order to make a quantitative statement about the validity of Newtonian theory we define two characteristic redshifts. Let $z_{\rm NL}$ be the redshift at which the perturbation on a given scale  goes nonlinear, which is, according to the spherical collapse model \cite{liddle}, given by
\begin{equation}
\mathcal{P}_{\delta}^{(1)}(k,z_{\rm NL})\sim 1,
\end{equation}
where $\mathcal{P}_{\delta}^{(1)}\equiv\frac{k^3}{2{\pi}^2}|\delta^{(1)}|^2$ is the dimensionless matter power spectrum. Perturbation theory can only be applied up to this redshift. For comparison, let $z_{\rm NN}$ be the redshift at which the relativistic modifications become as important as the second order Newtonian corrections, that is,
\begin{equation}
\delta^{(2)}(k,z_{\rm NN})\sim \delta_{H}(k,z_{\rm NN}). 
\end{equation}
In table \ref{anlann} we show an overview of these characteristic redshifts for different comoving scales, which can be read off from figure \ref{d1d2dk}, using the relation $a=1/(1+z)$ between scale factor and redshift. 
A discussion follows in the next section.

In figure \ref{dhvsk} we plot  the dimensionless band power of $\delta_H$ against the comoving wavenumber $k$ for different redshifts. It can be seen that the relativistic corrections to the Hubble rate on the cluster scale ($k\sim 0.1\,\rm Mpc^{-1}$) is of the order of 1\% at a redshift of $z\sim  2$ and 10\% today.

Note that relativistic corrections on the Gpc-scale are always below 1\%. However, this result should not be interpreted too fast as a confirmation of Newtonian theory on these very large scales. In fact, the density contrast itself is very small on the these scales. In particular, the fraction 
$\delta_H / \delta^{(1)}_{\rm N}=a^2 / 3 $ is independent of scale, so that the relativistic corrections become comparable to the density contrast today on all scales that still behave linear.

\section{Discussion}

The theory of structure formation gives the transition between the early homogeneous Universe and the Universe we observe today. In this work we have studied two different theories that describe this transition: relativistic and Newtonian cosmological perturbation theory. Although the former is the physically more correct tool, the latter is, due to its simplicity, the preferred choice in cosmological N-body simulations, with which we test our understanding of structure formation. Fixing the background universe to be an Einstein-de Sitter model, we have calculated the first and second  order perturbations in Newtonian cosmological perturbation theory and the first order perturbations in relativistic cosmological perturbation theory.

We have shown how the power spectra for matter and velocity perturbations behave according to relativistic cosmological perturbation theory in different gauges and according to Newtonian theory. In the synchronous/comoving gauge the density perturbations behave exactly like in Newtonian theory, however there are no velocity perturbations. In the longitudinal gauge the velocity perturbations behave like in Newtonian theory, but the density contrast differs significantly on superhorizon scales.

In the newly defined Newtonian matter gauge there are no relativistic corrections at all to the Newtonian matter- and velocity power spectra. However, there are other quantities present which do not appear in Newtonian cosmology: shear, intrinsic curvature and perturbations in the expansion rate. In particular, we have introduced the local modification of the expansion rate, $\delta_{H}$, and compared it to second order terms in Newtonian theory in order to evaluate at which scales Newtonian theory represents a good approximation.

Clearly, Newtonian theory is a good approximation when the scale goes nonlinear before relativistic modifications can become more important than second order Newtonian corrections,  i.e. for $z_{\rm NL}>z_{\rm NN}$. This is true for $k\gtrsim 0.6\,\rm{Mpc}^{-1}$, see table \ref{anlann}. For $z_{\rm NL}<z_{\rm NN}$, relativistic modifications become more important than second order Newtonian corrections \emph{before} the corresponding scale goes nonlinear. This is the case for $k\lesssim 0.6\,\rm{Mpc}^{-1}$.

Here, we only focussed on \emph{one} relativistic effect, the local variation of the Hubble parameter. There are however also other relativistic quantities that Newtonian simulations do not "see", namely intrinsic curvature and shear. When the fluctuations in the Hubble expansion rate become relevant, the tracing of light rays through a Newtonian simulation also must start to deviate from what the full theory would predict. This has to be kept in mind when using huge volume simulations in quantitative comparison with real data.

We stress that our results do not contradict the claims about the exact Newtonian-relativistic correspondence of equations of motions written in terms of Bardeen variables in different foliations found in literature, e.g.~by Hwang and Noh \cite{Hwang:2005,Hwang:2012}.  We have calculated the relativistic corrections as they appear in one specific and fixed foliation of space-time, i.e. as measured by one specific hypothetical observer in the linearly perturbed world, while the work by Hwang and Noh is based on the combination of gauge-invariant variables that can be interpreted on different hypersurfaces.

We are now in a position to be able to quantify the reliability of Newtonian simulations at large scales. We can interpret existing Newtonian simulations (i.e. simulations that have not been corrected or modified along the lines proposed in \cite{Chisari:2011,Green:2011}) to correspond to simulations in the Newtonian matter gauge. Then the simulated density contrast and peculiar velocities are correct (by definition) and other observables like the Hubble expansion rate are modeled very well on scales below 10 Mpc. However, on larger scales relativistic corrections are more important than Newtonian non-linear effects.

Thus we conclude, \emph{reliable predictions on supercluster scales, void scales, the baryon acoustic oscillation scale and beyond can only be based on general relativistic equations}. Newtonian simulations are good to provide us with a qualitative picture at scales above 10 Mpc, but a measurement of cosmological parameters at better than 10 per cent level cannot rely on them. Our findings do not question that Newtonian simulations remain to be important and extremely valuable at the cluster scale and below, i.e. in the deeply non-linear regime.

In Newtonian cosmology, the so-called redshift space distortions \cite{Samushia:2012iq} are due to the peculiar motion of galaxies, whereas in the relativistic formulation in Newtonian matter gauge, they would receive contributions from the peculiar motion and from space-time metric perturbations, most importantly the local variation of the cosmic expansion rate $\delta_H$. This clearly demonstrates that the differences between Newtonian and relativistic dust cosmologies should be observable in high fidelity galaxy redshift surveys. A quantitative analysis of these differences requires further investigations.

Let us finally remark that also exact theorems support our point of view that general relativistic and Newtonian dust cosmology are inequivalent. It has been shown by Ellis that there are no shear-free dust solutions of the Einstein field equations that both expand and rotate ("Dust Shear-Free Theorem", see \cite{Ellis:2011pi}). However, solutions of this kind do exist in Newtonian cosmology, as was shown by Narlikar \cite{Narlikar:1963}.  This alone demonstrates that the two theories are inequivalent.

\bigskip

\begin{acknowledgments}

We thank Julian Adamek, Stephen Green and Laura Lopez Honorez for useful discussions. We acknowledge support by Deutsche Forschungsgemeinschaft (DFG) via the International Research Training Group 881 `Quantum Fields and Strongly Interacting Matter' and Research Training Group 1620 `Models of Gravity'. 
\end{acknowledgments}

\bibliography{bib.bib}
\end{document}